\documentstyle[bezier,sprocl]{article}

\def\Journal#1#2#3#4{{#1} {\bf #2}, #3 (#4)}

\def\PLB{{\em Phys. Lett.}  B}

\def\ra{\rightarrow}
\def\be{\begin{equation}}
\def\ee{\end{equation}}
\def\bea{\begin{eqnarray}}
\def\eea{\end{eqnarray}}

\newcommand{\non}{\nonumber \\*}

\def\tr{\,{\rm tr}\,}
\def\e{{\,\rm e}\,}
\def\LA{\left\langle}
\def\RA{\right\rangle}
\newcommand{\rf}[1]{(\ref{#1})}
\newcommand{\eq}[1]{Eq.~(\ref{#1})}

\def\l{\lambda}

\def\eps{\epsilon}
\def\d{\delta}
\def\+{\dagger}
\def\df{\bar{F}}
\def\dw{\bar{W}}
\def\dww{\bar{W}W}
\def\ddww{\bar{W}\!\!\wedge\! W}
\def\db{B^{\+}}
\def\dbb{B^{\+}B}
\def\RAG{\right\rangle_{{\rm Gauss}}}
\def\u{\mbox{\boldmath $u$}}

\newcommand{\ie}{{\it i.e.}\ }

\def\L{{\cal L}}
\def\G{{\cal G}}

\begin{document}

\begin{flushright}
ITEP--TH--39/96 \\
hep-th/9608172\\
August, 1996
\end{flushright}
\vspace{12pt}

\begin{center}
{\bf APPLICATIONS OF SUPERSYMMETRIC MATRIX MODELS}
\footnote{Talk at the 2nd Int.\ Sakharov Conf.\ on Physics,
Moscow May 20--24, 1996}
\end{center}

\title{\mbox{} \vspace{-13pt} }

\author{Yu.\ MAKEENKO \vspace{3pt}}

\address{Institute of Theoretical and Experimental Physics, \\
B. Cheremushkinskaya 25, 117259 Moscow, Russia\\
{\rm and} \\
The Niels Bohr Institute,\\
Blegdamsvej 17, DK-2100 Copenhagen \O, Denmark
}

\maketitle\abstracts{
Matrix models have wide applications in nuclear theory,
condensed matter theory and quantum field theory. I discuss supersymmetric
extensions of matrix models and their applications to
branched polymers, the meander problem, and superstrings in lower
dimensions.}

\section{Introduction}

Matrix models have wide applications in nuclear theory,
condensed matter theory, high energy theory and quantum gravity
since the seminal paper by Wigner~\cite{Wig51}.  An extension to
supermatrices can be found in Refs.~\cite{VWZ85,Zuk94,AM91}.

I report in this talk some results on another supersymmetric
extension~\cite{MP95,AMZ96} of matrix models, which is based on one
complex bosonic matrix $B$ and one fermionic matrix $F$ and
is along the line proposed by Marinari and Parisi~\cite{MP90}.
I discuss applications of the supersymmetric matrix models
to branched polymers,
the meander problem, and superstrings in lower dimensions.

\section{Supersymmetric matrix models}

The supersymmetric matrix
models in the $D=0$ dimensional target space  are built out of
the ``superfields''~\cite{MP95}
\be
W_a= \left(B, F\right),~~~~~~\bar{W}_a= \left(B^\dagger, \bar{F}\right),
\label{superfields}
\ee
where $a=1,2$ while $B$ and $F$ are general complex bosonic and
fermionic (\ie Grassmann valued) $N\times N$ matrices, respectively.
In other words, the hermitean conjugated $B^\dagger \neq B $
and the Grassmann involuted $\bar F \neq F $.

There is no need of introducing a superspace coordinate $\theta$
which would be dimensionless in $D=0$,
since the propagators for both bosonic and fermionic matrices coincide:
\be
\LA  B_{ij} B^\dagger_{kl}\RAG = \frac 1N \delta_{il} \delta_{kj} \,,
~~~~~~ \LA  F_{ij} \bar{F}_{kl}\RAG = \frac 1N \delta_{il} \delta_{kj} \,.
\label{BFpropagators}
\ee
Hence,
the supersymmetry reduces in $D=0$ simply to rotations between the
$B$- and $F$-compo\-nents. The proper transformation reads
\bea
\d_{\eps}B^{\+}=\df\eps\,,& &~~~~\d_{\eps}F=-\eps B\,,
\label{susy1} \\
\d_{\bar{\eps}}B=\bar{\eps}F\,,& &~~~~\d_{\bar{\eps}}
\df=-\db\bar{\eps}\,,
\label{susy2}
\eea
where $\epsilon$ and  $\bar\epsilon$ are Grassmann valued.
Note that it is a huge symmetry
since the parameters $\eps$ and $\bar \epsilon$ are $N\times N$
matrices.

The simplest Gaussian supersymmetric potential reads
\be
V_{\rm Gauss}=N \tr \bar{W} W,
\label{SVGauss}
\ee
where
\be
\bar{W} W\equiv\sum_{a=1}^2 \bar{W}_a W_a =
  B^\dagger B +\bar{F} F,
\label{barWW}
\ee
which reproduces the propagators~\rf{BFpropagators}.
It is obviously invariant under the rotation~\rf{susy1}--\rf{susy2}.
It is also clear from \eq{SVGauss} why one needs complex matrices
in $D=0$: the trace of the square of a fermionic matrix vanishes.

Any potential, which is symmetrically
constructed from the ``superfields'' \rf{superfields}, is
supersymmetric so that contributions from the loops of the bosonic
and fermionic matrix fields are mutually cancelled which is the key
property of the supersymmetry.

A general interaction potential, which is invariant under the matrix
supersymmetry transformation~\rf{susy1}--\rf{susy2}, reads
\be
V_{\rm gen}\left( \bar{W} W\right)
=N \sum_{k\geq 1} \frac{g_k}{k}\tr \left( \bar{W} W\right)^k,
\label{Vgen}
\ee
where $g_k$ are the coupling constants.
This invariance can be seen from
\be
\delta_\epsilon \dww = \delta_\epsilon \left(
  B^\dagger B +\bar{F} F\right)
 = \df \eps B -\df \eps B =0.
\ee
The supersymmetric matrix model with the potential~\rf{Vgen}
describes branched polymers as is shown in the next Section.

The supersymmetry transformations~\rf{susy1}, \rf{susy2}
can be formalized by introducing the (matrix) supercharges
\footnote{Here and below the order reflects matrix multiplication.}
\be
Q_{ij}=\sum_{k=1}^N \left( F_{ik}\frac{\partial}{\partial B_{jk}}
-\frac{\partial}{\partial \bar{F}_{ki}} B^\dagger_{kj}\right),~~~~
\bar{Q}_{ij}=\sum_{k=1}^N \left( \frac{\partial}{\partial
{B}^\dagger_{ki}}
\bar{F}_{kj} - B_{ik}
\frac{\partial}{\partial F_{jk}} \right),
\label{defQij}
\ee
so that
\be
\delta_{\epsilon}\ldots=
\left[ \tr \bar{Q} \epsilon,\,\ldots\,\right],~~~~~~~
\delta_{\bar\epsilon}\ldots=
\left[ \tr \bar{\epsilon} Q ,\,\ldots\,\right].
\ee
Their commutators read
\bea
\left\{ Q_{ij}\,,\;Q_{mn} \right\}&=&
\left\{ \bar{Q}_{ij}\,,\;\bar{Q}_{mn} \right\}=0, \\
\left\{ {Q}_{ij}\,,\;\bar{Q}_{mn} \right\}&=&
-\delta_{in}
\sum_{k=1}^N \left( B_{mk}\frac{\partial}{\partial {B}_{jk}}
+\frac{\partial}{\partial {B}^\dagger_{km} } B^\dagger_{kj}
\right) \non & &
- \sum_{k=1}^N \left( F_{ik}\frac{\partial}{\partial {F}_{nk}}
-\frac{\partial}{\partial \bar{F}_{ki} } \bar{F}_{kn}
\right)\delta_{mj}.
\label{QQbar}
\eea

\section{Application to branched polymers}

The partition function of the supersymmetric matrix model with the
potential~\rf{Vgen} reads
\be
Z\left[ g \right]\equiv\int \prod_{a,b=1}^2 d W_a d \dw_b
\e^{-V_{\rm gen}\left( \bar{W} W\right)} =1 .
\label{BPpartition}
\ee
It is equal to $1$ because of the cancellation of bosonic and
fermionic loops due to the supersymmetry.

Likewise, all the supersymmetric correlators vanish, for example
\be
\LA \frac 1N \tr \left( \dww \right)^n \RA \equiv
\int dW d \dw
\e^{-V_{\rm gen}\left( \bar{W} W\right)}
\frac 1N \tr \left( \dww \right)^n =0 .
\ee

Physical quantities of the model are described
by the correlators of the pure bosonic matrices, which are
nontrivial. Their generating function is
\begin{equation}\label{loop}
G(\l)=\left\langle\frac{1}{N}\tr\frac{1}{\l-\db B}\right\rangle
=\frac{1}{\l}+\sum_{n=1}^{\infty}\frac{1}{\lambda^{n+1}}G_n
\end{equation}
with
\be
G_n\equiv\left\langle\frac{1}{N}\tr(\dbb)^n\right\rangle .
\label{defGn}
\ee
The imaginary part of $G(\l)$ determines the distribution of
eigenvalues of the matrix $B^\dagger B$ and, therefore, the
spectrum of the proper statistical model.

The correlators~\rf{defGn} can be calculated~\cite{AMZ96} using
the Schwinger--Dyson equations and the supersymmetry Ward
identities. As a result, $G(\l)$ obeys at large $N$ the closed
equation
 \begin{equation}\label{f'}
 \left(\lambda G(\lambda)-1\right)V'\left(\lambda-{1}/{G(\lambda)}
 \right)=\lambda G^2(\lambda)\,,
 \end{equation}
whose limit of $\l\ra\infty$ yields the equation
\begin{equation}\label{f1}
 G_1V'(G_1)=1
\end{equation}
for $G_1$. This equation is quadratic for a quartic potential.

\begin{figure}[t]
\unitlength=0.70mm
\linethickness{0.4pt}
\begin{picture}(153.00,50.00)(0,10)
\bezier{124}(20.00,15.00)(6.00,24.00)(15.00,35.00)
\bezier{48}(15.00,35.00)(20.00,38.00)(25.00,35.00)
\bezier{128}(25.00,35.00)(35.00,25.00)(20.00,15.00)
\bezier{104}(50.00,15.00)(37.00,21.00)(45.00,30.00)
\bezier{52}(45.00,30.00)(50.00,34.00)(55.00,30.00)
\bezier{100}(55.00,30.00)(62.00,21.00)(50.00,15.00)
\bezier{108}(75.00,15.00)(63.00,22.00)(74.00,30.00)
\bezier{124}(76.00,30.00)(89.00,22.00)(75.00,15.00)
\bezier{76}(74.00,30.00)(63.00,33.00)(65.00,40.00)
\bezier{16}(65.00,40.00)(66.00,42.00)(68.00,42.00)
\bezier{72}(68.00,42.00)(73.00,43.00)(75.00,30.00)
\bezier{48}(75.00,30.00)(77.00,40.00)(79.00,41.00)
\bezier{72}(76.00,30.00)(87.00,33.00)(86.00,40.00)
\bezier{40}(86.00,40.00)(84.00,44.00)(79.00,41.00)
\bezier{108}(105.00,15.00)(93.00,22.00)(104.00,30.00)
\bezier{116}(105.00,15.00)(118.00,21.00)(106.00,30.00)
\bezier{108}(104.00,30.00)(93.00,38.00)(104.00,45.00)
\bezier{108}(106.00,30.00)(117.00,37.00)(106.00,45.00)
\bezier{76}(104.00,45.00)(93.00,48.00)(95.00,55.00)
\bezier{16}(95.00,55.00)(96.00,57.00)(98.00,57.00)
\bezier{72}(98.00,57.00)(103.00,58.00)(105.00,45.00)
\bezier{48}(105.00,45.00)(107.00,55.00)(109.00,56.00)
\bezier{72}(106.00,45.00)(117.00,48.00)(116.00,55.00)
\bezier{40}(116.00,55.00)(114.00,59.00)(109.00,56.00)
\bezier{100}(135.00,15.00)(123.00,19.00)(128.00,30.00)
\bezier{100}(135.00,15.00)(147.00,19.00)(142.00,30.00)
\bezier{84}(128.00,30.00)(116.00,32.00)(119.00,40.00)
\bezier{28}(119.00,40.00)(120.00,43.00)(124.00,43.00)
\bezier{80}(123.00,43.00)(131.00,43.00)(129.00,31.00)
\bezier{68}(129.00,31.00)(135.00,37.00)(141.00,31.00)
\bezier{76}(142.00,30.00)(153.00,32.00)(151.00,40.00)
\bezier{28}(151.00,40.00)(149.00,43.00)(146.00,43.00)
\bezier{80}(147.00,43.00)(138.00,42.00)(141.00,31.00)
\put(20.00,15.00){\circle*{2.00}}
\put(50.00,15.00){\circle*{2.00}}
\put(75.00,15.00){\circle*{2.00}}
\put(105.00,15.00){\circle*{2.00}}
\put(135.00,15.00){\circle*{2.00}}
\put(35.00,22.00){\makebox(0,0)[cc]{=}}
\put(63.00,22.00){\makebox(0,0)[cc]{+}}
\put(90.00,22.00){\makebox(0,0)[cc]{+}}
\put(118.00,22.00){\makebox(0,0)[cc]{+}}
\put(152.00,22.00){\makebox(0,0)[cc]{+ $\cdots$}}
\put(14.00,22.00){\line(1,1){11.03}}
\put(13.00,24.00){\line(1,1){10.99}}
\put(13.00,27.00){\line(1,1){8.00}}
\put(14.00,31.00){\line(1,1){4.01}}
\put(15.00,20.00){\line(1,1){11.00}}
\put(17.00,19.00){\line(1,1){9.98}}
\put(19.00,18.00){\line(1,1){7.98}}
\put(21.00,17.00){\line(1,1){5.00}}
\end{picture}
\begin{picture}(153.87,62.21)(-3,9)
\bezier{372}(10.00,10.00)(17.00,58.00)(30.00,15.00)
\bezier{372}(50.00,10.00)(57.00,58.00)(70.00,15.00)
\bezier{372}(90.00,10.00)(97.00,58.00)(110.00,15.00)
\bezier{372}(130.00,10.00)(137.00,58.00)(150.00,15.00)
\bezier{52}(59.02,35.31)(54.36,40.30)(50.03,45.04)
\bezier{52}(59.02,35.31)(63.92,40.38)(68.01,45.04)
\put(68.01,45.04){\circle{1.00}}
\put(50.03,45.04){\circle{1.00}}
\put(69.97,15.03){\circle{1.00}}
\put(59.02,35.31){\circle*{1.00}}
\bezier{204}(99.04,35.36)(102.91,62.21)(109.97,39.10)
\bezier{48}(104.02,49.89)(100.14,54.32)(96.96,59.03)
\bezier{44}(104.02,49.89)(107.62,54.32)(110.94,59.03)
\bezier{52}(134.32,29.97)(130.17,34.81)(126.02,39.93)
\bezier{52}(144.42,29.97)(149.13,34.81)(153.00,39.93)
\put(96.96,59.03){\circle{1.00}}
\put(110.94,59.03){\circle{1.00}}
\put(126.02,39.93){\circle{1.00}}
\put(153.00,39.93){\circle{1.00}}
\put(149.96,15.03){\circle{1.00}}
\put(144.42,29.97){\circle*{1.00}}
\put(134.46,29.97){\circle*{1.00}}
\put(99.04,35.36){\circle*{1.00}}
\put(109.97,15.03){\circle{1.00}}
\put(104.02,49.75){\circle*{1.00}}
\put(109.97,39.10){\circle{1.00}}
\put(30.00,15.00){\circle{1.00}}
\put(10.00,10.00){\circle*{2.00}}
\put(50.00,10.00){\circle*{2.00}}
\put(90.00,10.00){\circle*{2.00}}
\put(130.00,10.00){\circle*{2.00}}
\end{picture}
\caption {
   Typical diagrams for $G_1$
   when the potential $V(\dww)$ is cubic in $\dww$ (above)
and the associated branched polymer graphs (below).}
 \label{cactus}
\end{figure}
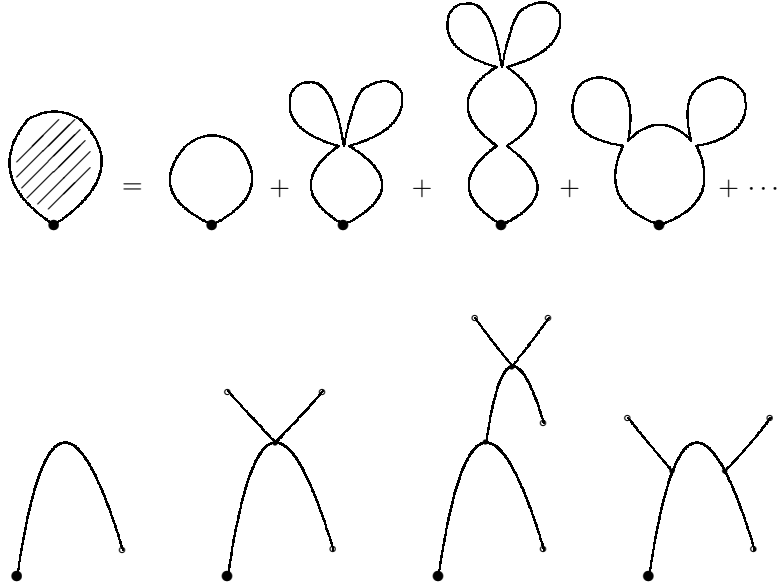

The fact that
a closed equation is obtained for the propagator $G_1$ is a
consequence of the cancellations between bosonic and fermionic loops.
Some of the diagrams which survive
the cancellation are shown in Fig.~\ref{cactus}.
For obvious reasons, they are called
the ``cactus diagrams''.
Note that the diagrams have orientation: the cactus loops can only
proliferate on the exterior of already existing loops. This is in
contradistinction to related bubble diagrams one encounters in the
large-$N$ limit of pure bosonic or fermionic vector models.

The critical behavior of the model arises
when \eq{f1}  holds simultaneously with
 \begin{equation}\label{dec}
 \left[G_cV'(G_c)\right]'
 =\ldots=
 \left[G_cV'(G_c)\right]^{(m-1)}=0,
 \end{equation}
 which can always be achieved by tuning $m$ couplings $g_k$
 of the potential~\rf{Vgen}.

 Near the critical point, $G_1$ behaves as
 \begin{equation}\label{f1c}
 G_1\simeq G_c- {\rm const}\cdot(\alpha_c-\alpha)^{1/m},
 \end{equation}
 where $\alpha$ stands for an overall scale of $g_k$'s.
The susceptibility
 $\chi\equiv {\partial G_1}/{\partial \alpha}$ at the critical
 point scales as
 \be
 \chi\sim(\alpha_c-\alpha)^{-\gamma_{\rm str}}
 \ee
 with $ \gamma_{\rm str}=1-{1}/{m}$,
 which coincides with the (multi-)critical
 index of the branched polymers~\cite{polymer}.

This relation to branched polymers becomes explicit by noting that
the cactus graphs which survive the supersymmetric cancellation
have an interpretation as branched polymers,
with the couplings $-g_k$ associated with the branching weights.
Cutting each loop in a succession,
such that
only the stating line is attached to the vertex, produces a branched
polymer graph of a ``chiral'' type since  branching only occurs
at one side of the open line, corresponding to the fact that
the cactus loops can only be attached to the exterior of already
existing loops.  This is illustrated in Fig.~\ref{cactus}.
Equation~\rf{f1} can then be rederived pure combinatorially.

\section{Application to the meander problem}

The meander problem is to calculate combinatorial numbers associated with
the crossings of an infinite river (Meander) and a closed road
by $2n$ bridges.%
\footnote{See Ref.~\cite{DGG95} for an introduction to the subject.}
Neither the river nor the road intersects with itself.
These meander numbers, $M_n$, obviously describe the number of
different foldings of a closed strip of $2n$ stamps or of a closed
polymer chain.

The generating function of the meander numbers can be represented
via the following correlator in the supersymmetric matrix
model~\cite{MP95}:
\be
M(c)\equiv \sum_{n=1}^\infty c^{2n}  M_n = \lim_{N\ra\infty}
\LA \frac 1N \tr{B B^\dagger}
\ln{\left(\int d\phi_1 d\phi_2 \e^{-S}\right)}\RAG \,,
\label{Sm}
\ee
where $\phi_1$ and $\phi_2$ are $N\times N$ hermitean matrices,
the action $S$ is given by
\be
S=\frac N2 \tr{\phi^2_1}+\frac N2 \tr{\phi^2_2}
-cN \tr{\left(\phi_1 B^\dagger \phi_2 B\right)}
-cN \tr{\left(\phi_1 \bar F \phi_2 F\right)}
\label{Saction}
\ee
and the Gaussian averaging is with respect to the action~\rf{SVGauss}.
The presence of the log in \eq{Sm}
leaves only one loop of the field $\phi$ associated with
the road while the supersymmetry kills the loops of
the field $W$ associated with the river. The limit of
large $N$ is needed to keep only the planar graphs as in
the original meander problem.

Expanding in $c$, the coupling constant of the quartic interaction,
the meander numbers can be represented as the sum over words built
out of two letters:
\bea
M_n &=& \sum_{a_2,\cdots, a_{2n-1}, a_{2n}=1}^2
\LA \frac 1N \tr{B \bar{W}_{a_2} \cdots
W_{a_{2n-1}} \bar{W}_{a_{2n}}}
\RAG   \non & & ~~~~~~~~~~\times
\LA \frac 1N \tr{\bar{W}_{a_{2n}}
W_{a_{2n-1}} \cdots  \bar{W}_{a_{2}} B}\RAG\,,
\label{Swords}
\eea
where the order of matrices is
essential for the fermionic components.
Equation~\rf{Swords} is a nice representation of
the meander numbers which looks more natural than the one based on the
replica trick in a pure bosonic model.

Equation~\rf{Swords} can be represented in an alternative form by
introducing noncommutative variables $u,v$ and $u^\dagger,v^\dagger$
which
are annihilation and creation operators in a Hilbert space with the
vacuum $\left|\Omega \RA$
and obey the Cuntz algebra
\be
uu^\dagger=1,~~vv^\dagger=1,~~
uv^\dagger=0,~~vu^\dagger=0,
\ee
as well as
the completeness condition
\be
u^\dagger u + v^\dagger v =1 - \left|\Omega \RA \LA \Omega \right| \,.
\label{completeness}
\ee
There are no more relations between the noncommutative variables.

Denoting
\be
\u_a=(u,v),~~~~~~\bar{\u}_a=(u,-v),
\ee
the generating function~\rf{Sm} can alternatively be represented
as the vacuum expectation value
\be
M(c) =  \LA\Omega | \bar{G} u^\dagger u {G}^\dagger
 |\Omega \RA = - \LA\Omega | \bar{G} v^\dagger v {G}^\dagger
 |\Omega \RA,
\label{barsmea}
\ee
where $G$ is given by the continued fraction
\be
G\left(\u \right)=\frac{1}{\displaystyle 1-\sqrt{c}\u_{a_1}\frac{1}
{\displaystyle 1-\sqrt{c}\bar{\u}_{a_2}
\frac{1}{\displaystyle 1
-\sqrt{c}\u_{a_3}\frac{1}{\displaystyle
1-\sqrt{c}\bar{\u}_{a_4} \frac{1}{\vdots}\,
\bar{\u}_{a_4}}\,\u_{a_3}}\,\bar{\u}_{a_2}}\,\u_{a_1}} .
\label{SCvi}
\ee
Here $\u$ and $\bar{\u}$ interchange in the consequent lines.
The following notations are used in \eq{barsmea}:
\be
G\equiv G\left(\u \right),~~
\bar{G}\equiv G\left(\bar{\u} \right),~~
G^\dagger\equiv G\left(\u^\dagger \right),~~
\bar{G}^\dagger\equiv G\left(\bar{\u}^\dagger \right).
\ee
The two expressions on the right hand side of
\eq{barsmea} are equal due
to the supersymmetry.

Though \eq{barsmea} reproduces the
meander numbers when expanded in $c$, it seems to look like
a reformulation rather than
a solution to the problem since it is not clear how
to deal with functions of noncommutative variables.

\section{Application to superstrings?}

The critical index of the
string susceptibility $\gamma_{\rm str}$ for
a superstring embedded in a $D$-dimensional space had been calculated
from the super-Liouville theory and reads~\cite{PZ88}
\be
\gamma_{\rm str}= \frac{D-1-\sqrt{(1-D)(9-D)}}{4}.
\label{PZformula}
\ee

Many (not yet successful) attempts of discretizing superstring
are performed starting from~\cite{MS90}.
A progress has been achieved~\cite{AG92}
only for the simplest case of pure 2-dimensional supergravity
which can be associated with a supereigenvalue model.
It reveals the super-Virasoro algebra associated with
the Neveu--Schwarz sector of the superstring.


The idea to verify whether or not a super-Virasoro algebra can be
realized in the supermatrix models is to construct the matrix
generators (cf.~\rf{defQij}, \rf{QQbar})
\be
\L_{ij}=\sum_{k=1}^N \left( \frac{\partial}{\partial B_{ki}} B_{kj}
-\frac{\partial}{\partial F_{ki}} F_{kj}\right),~~~~
\G_{ij}=\sum_{k=1}^N \left( \frac{\partial}{\partial B_{ki}} F_{kj}
+\frac{\partial}{\partial F_{ki}} B_{kj}\right)
\label{defGij}.
\ee
Here $\L_{ij}$ is Grassmann even and $\G_{ij}$ is Grassmann odd.

The operators  $\L_{ij}$ and $\G_{ij}$ obey the commutation relations
\bea
\left[\L_{ij}\,,\;\L_{mn} \right]& =& \delta_{in} \L_{mj} -
 \L_{in} \delta_{mj} \,,
\label{LL} \\
\left[\G_{ij}\,,\;\L_{mn} \right]& = &\delta_{in} \G_{mj}
 -  \G_{in} \delta_{mj}
 \label{GL} \,, \\
\left\{\G_{ij}\,,\;\G_{mn} \right\} &=& \delta_{in} \L_{mj} +
 \L_{in} \delta_{mj} \,.
\label{GG}
\eea
These can be derived by explicitly commuting the operators 
\rf{defGij}.

The commutator~\rf{LL} itself implies the Virasoro algebra
\be
\left[ \L_s,\,\L_t \right]=(t-s)\L_{s+t}\,,~~~~~~
\L_s=\tr\left( \L \left(\dww \right)^s\right)
\label{Virasoro}
\ee
for $s,t \geq 0$ as $N\ra\infty$. Given~\rf{Virasoro},
the potential~\rf{Vgen} can then be recovered from
the Virasoro constraints
\be
0=\int dW d \dw \L_s
\e^{-V_{\rm gen}\left( \bar{W} W\right)}=
L_s\left[g\right] Z\left[g\right] .
\ee

The conjecture is that the whole matrix algebra~\rf{LL}--\rf{GG}
implies, as \mbox{$N\ra\infty$},
the super-Virasoro algebra associated with the Ramond
sector of the superstring in the $D=0$ dimensional target space:
\be
\left[ \L_s,\,\L_t \right]=(t-s)\L_{s+t}\,,~~~~
\left[ \G_r,\,\L_s \right]=\left(\frac s2-r\right)\G_{r+s}\,,~~~~
\left\{ \G_r,\,\G_s \right\}=2\L_{r+s}\,.
\label{super-Virasoro}
\ee
An explicit form of the operators $\L_s$ and $\G_r$
can be constructed starting from
\be
\L_0=\tr \L,~~~~\G_0=\tr \G,~~~~
\L_1=a\tr \L \dww +b \tr \G \ddww
\ee
where
\be
\ddww\equiv
\sum_{a,b=1}^2 e^{ab} \bar{W}_a W_b = B^\dagger F - \bar F B
\label{wedgeWW}
\ee
and using~\rf{super-Virasoro}. For example, one gets
\bea
\G_1&=&c\tr \G \dww +d \tr \L \ddww, \non
\L_2&=&c(c+d)\tr \L \left(\dww\right)^2 + c(c-d) \tr \G \ddww \dww
\eea
with
\be
c=-2b,~~~~~~d=2a+4b
\ee
and so on. The constant $b\neq 0$ in $\L_1$ since $\L_2$
vanishes otherwise. In order for this procedure to be
successful, all the operators $\L_s$ and $\G_r$ are to
be nonvanishing.

The action of the proper supersymmetric matrix model should
then be determined to reproduce these super-Virasoro operators.
It should involve both $\dww$ and $\ddww$ given respectively by
Eqs.~\rf{barWW} and \rf{wedgeWW} and, therefore, Grassmann
odd coupling constants in addition to those in \rf{Vgen}.

\section*{Acknowledgments}
This work was supported in part by the grant INTAS--94--0840.

\end{document}